\documentstyle[12pt,fleqn]{article}
\begin{document}
\baselineskip= 22 truept
\newcommand{\nc}{\newcommand}
\begin{flushright}
IP/BBSR/96-11\\
hep-th/ 9602145\\
\end{flushright}
\vspace{1cm}\begin{center} 
{\large \bf K-models and Type IIB Superstring Backgrounds}\\
\vspace{1cm} 
{\bf {\sc 
Alok Kumar}$^1$ } and {\bf {\sc Gautam Sengupta}$^2$}\\
\vspace{0.5cm}
$^1$Institute of Physics\\
Bhubaneswar 751 005, INDIA\\
e-mail: kumar@iopb.ernet.in\\
\bigskip
$^{2,*}$Indodeep Housing Complex\\
256/5, Dum Dum Road\\
Calcutta 700 074, INDIA.
\end{center} 
\thispagestyle{empty}
\vskip 2cm
\begin{abstract}
A family of type IIB superstring backgrounds 
involving Ramond-Ramond
fields are obtained in ten dimensions starting from a K-model
through a generalization of our recent results.
The unbroken global $SL(2,R)$ symmetry of the type IIB 
equations of motion are implemented in this context as a solution
generating transfromation. A geometrical analysis, based on the 
tensor structure of the higher order $\alpha^{\prime}$ terms in the 
equations of motion, is employed to show that these backgrounds are
exact.
\end{abstract}
\vfil
\hrule
\leftline {$^*$Residential address.}
\eject
\nc {\be}{\begin{equation}}
\nc {\ee}{\end{equation}}
\nc {\bea}{\begin{eqnarray}}
\nc {\eea}{\end{eqnarray}}
\nc {\gam} {\Gamma}
\nc {\hi}{H_{\lambda\mu\nu}}
\nc {\pa}{\partial}
\nc {\rc}{R_{\lambda\mu\nu\kappa}=
2l_{[\lambda}\pa_{\mu ]}\pa_{[\nu}Fl_{\kappa ]}}
\nc {\rci}{R_{\lambda\mu\nu\kappa}}
\nc {\ric}{R_{\mu\nu}}
\nc {\hf}{{1\over 2}}
\nc {\pab}{{\bar \partial}}
\noindent {\bf 1. Introduction}
\bigskip

\noindent Recent advances in string theory
has provided a glimpse of their underlying
non-perturbative structure. There now seems to be strong
indication that the several distinct string theories 
all arise from a yet
unknown fundamental theory in eleven dimensions,
which has been christened
as M-theory \cite{mref}. 
The central guiding principle behind this unification has been
the duality symmetries \cite{tdual,sdual,senrev,jhs}. 
They include the T-duality \cite{tdual} which is perturbative
in the genus expansion and the conjectured 
non-perturbative S-duality \cite
{sdual,senrev,jhs}. The latter relates
electrical charged perturbative states to 
magnetic charged solitons and weak to strong coupling regimes.
As an example, we have the type IIB string theory in ten dimensions
\cite{hull,berg}. These
possess a global $SL(2,R)$ symmetry of the 
classical effective field theory
which is broken to the discrete $SL(2,Z)$ S-duality group in the 
quantum theory which relates NS-NS to the R-R sector.

In the sigma model approach to string theory
\cite{sigmarev}, one considers the propagation
of a string in the background of its massless excitations. 
The evolution
is described by a two dimensional sigma model 
in which the background fields appear as couplings.
Conformal invariance then requires the corresponding perturbative 
beta functions for these couplings
to vanish, leading to the background field 
equations of motion. The higher
order terms \cite{sigmarev} in these equations provide the
stringy corrections to the
tree level theory and may be computed perturbatively. 
However for the type IIB string background, 
this remains a non-trivial exercise owing to the 
presence of R-R fields
which arise from the solitonic sector \cite{fms}. 
These couple to the spin fields
on the worldsheet and makes the corresponding perturbation theory
intractable at higher orders \cite{fms,pol}.
There exists however, a large class of string 
backgrounds for the bosonic and heterotic cases, 
where the higher order
contributions are identically zero 
\cite{tset1,berg1,tsetk,tset2,tsetrev}. 
Hence the tree level equations of motion are
exact to all orders in the sigma model coupling $\alpha^{\prime}$.
Amongst these are the class of backgrounds with a 
covariantly constant null 
Killing vector which are known as K-models \cite{tset1,tsetk}. 
The simplest example in this class are the
plane wave string backgrounds. 
For the bosonic and heterotic versions of these,
it has been shown that they are exact 
(in $\alpha^{\prime}$) through a purely
geometrical analysis based on the existence of the 
covariantly constant null
killing vector \cite{berg1,horo,duval,guv}.

Previously, in collaboration with Kar 
\cite{asg}, we have shown that
starting from such a plane wave background embedded trivially in a 
type IIB string theory (i.e with vanishing R-R fields),
it is possible to generate a non-trivial
type IIB background with R-R fields in ten dimensions. 
We further showed, through
a geometrical analysis, that these type IIB backgrounds were
also exact to all orders in $\alpha^{\prime}$. This method
avoids the complications arising from the 
worldsheet couplings of the R-R fields. In this article,
we extend our analysis to a more general class of K-models
with chiral couplings on the worldsheet\cite{tsetrev}. 
On compactification to lower dimensions,
these couplings lead to background gauge fields.
The bosonic and the heterotic versions of these K-models describe 
strings propagating in a uniform magnetic 
field background \cite{russ} as one of the special case. 
These can be formulated as exact
confornal field theories and illustrates a phase transition
at some critical value of the magnetic field where infinite
number of states becomes tachyonic.
In lower dimensions they also describe charged black holes
through the Kaluza-Klein mechanism \cite{tsetrev}.
We consider such a K-model in ten dimensions 
trivially embedded ( i.e without R-R fields ) in a 
type IIB background. Using the
global $SL(2,R)$ symmetry of the equations of 
motion, we generate a type
IIB background with non-trivial R-R field background. Subsequently,
assuming a specific structure for the field strengths, 
we show that the 
backgrounds obtained are exact (in $\alpha^{\prime}$) through a 
geometrical analysis. In this context, unlike our earlier work, we 
show that the geometrical considerations based on the 
covariantly constant null Killing
vector may be retranslated in the language of the 
index structure of the
corresponding higher order tensors\cite{duval,asg}. 
It is possible that the present approach 
is applicable to a wider class of models. 
This article is divided into four sections.
In Section-2 we present a brief review of  
the backgrounds obtained from K-models for the bosonic case
and show that they are exact. Section-3 deals 
with K-models embedded
in a type IIB theory and their $SL(2,R)$ transformations. In this
section we also explicitly prove that these backgrounds 
are exact in the presence of
R-R fields. We present the conclusions in section-4.
\bigskip

\noindent {\bf 2. String Backgrounds from K-Models.}

\noindent We begin with a description of the string background fields 
obtained from the K-models. The Lagrangian for the 
bosonic K-model is given as\cite{tsetrev}:
\bea
{\cal L}&=&   2\pa u \pab v 
+ K(u,x)\pa u \pab u + 2 A_{i}(u,x) \pa u 
\pab x^i
+ 2 {\bar A}_i(u,x) \pab u \pa x^i \\ \nonumber
&+& (G_{ij} + B_{ij})(u,x) \pa x^i \pa x^j
+R^{(2)}\phi (u,x). \label{kact}
\eea
We specialize to the case where $G_{ij}=\delta _{ij}$,
$B_{ij}=0$ and $\phi=\phi (u)$.
For this case we have the metric
\be
ds^2=2dudv + 2A_{i}^{+}(u,x)dudx^i 
+ K(u,x) du^2 + dx^idx_i \label{kmet},
\ee
and the antisymmetric tensor field:
\be
B_{\mu \nu}=\left ( \begin{array}{ccc}
0 & 1 & A_{i}^{-}\\
-1 & 0 & 0\\
-A_{i}^{-} & 0 & 0\\
\end{array} \right ). \label{kb}
\ee
The greek indices ($\mu, \nu $) run over $(0....9)$, 
$(u,v)$ are the light-cone coordinates and 
the latin indices $(i,j) = (2...9)$ run over the transverse space
coordinates $x^i$. We also have,
\be
A^{\pm}_{i}=A_{i}\pm {\bar A}_{i}. \label{ai}
\ee
The $v$ independence of the metric leads to a killing
vector $l^\mu = (0,1,0...0)$. It is possible to express the metric in a
compact form in terms of $l^\mu$ which eases subsequent computations.
Explicitly, we have
\be
G_{\mu\nu}=M_{\mu\nu} + K l_{\mu}l_{\nu} \label{gm},
\ee
where $M_{\mu\nu}$ is a $10 \times 10$ symmetric matrix 
\be
M_{\mu\nu}=\left (\begin{array}{ccc}
0 & 1 & A_i^+\\
1 & 0 & 0\\
A^+_i & 0 & I_8\\
\end{array} \right ),
\label{m}
\ee
and $I_{8}$ is a $8 \times 8$ unit matrix. 
The inverse metric is obtained as,
\be
G^{\mu\nu}=\left (\begin{array}{ccc}
0 & 1 & 0\\
1 & -K+A_{i}^{+2} & -A^+_i\\
0 & -A_{i}^{+} & I_{8}\\
\end{array} \right). \label{ginv}
\ee
The only non-zero connections are
$\gam^{i}_{uu}$, $\gam^{v}_{uu}$, $\gam^{v}_{ui}$, $\gam^{j}_{ui}$ and
$\gam^{v}_{ij}$. Using these connections, it is easy to show that the
null killing vector is covariantly constant. 
We therefore have $D_{\mu}l^{\nu}=0$
and $D_{\mu}l_{\nu}=0$. The curvature tensor for the backgrounds
(\ref{ginv}) may be obtained as
\be
\rci=\rci^{(M)} + 2l_{[\lambda}\pa_{\mu ]}\pa_{[\nu}Kl_{\kappa
]}. \label{rci}
\ee
Notice that the only non-zero independent components of 
$\rci$ are $R_{uiuj}$ and 
$R_{uijk}$. These are as follows:
\bea
R_{uiuj}& =& \hf \pa_i\pa_j K - \hf \pa_{u}\big [ \pa_iA_j^+ + 
\pa_jA_i^+ \big ]       \nonumber\\  
&-& {1\over{4}}G^{mn}\big ( \pa_jA_m^+ - \pa_mA_j^+\big )
\big ( \pa_iA_n^+ - \pa_nA_i^+\big ),  \label{rui}
\eea
and
\be
R_{uijk}=\hf \pa_i \big [ \pa_kA_j^+ - \pa_jA_k^+\big ]. \label{rul}
\ee
We get the expressions for the corresponding
components of the Ricci tensor 
by appropriate contractions 
of the Riemann tensor as,
\be
R_{uu}=\hf \pa^i\pa_iK -\pa_u\pa^iA_i^+ 
- {1\over {4}}(F_{jm} + {\bar{F}}_{jm})
(F^{jm} + {\bar{F}}^{jm}) \label{ru},
\ee
\be
R_{uk}=-\hf \pa^m(F_{km} + {\bar F}_{km}), \label{rk}
\ee
where
\be 
F_{ij}=\big ( \pa_iA_j- \pa_jA_i\big ),\;\;\;\;
{\bar{F}}_{ij}=\big ( \pa_i{\bar{A}}_j - \pa_j{\bar{A}}_i\big )
\label{fbar}
\ee
are the {\it field strengths} associated with the couplings
$A$ and $\bar{A}$.
Next we consider the antisymmetric tensor field strength which is given 
in the standard form as:
\be
H_{\lambda\mu\nu}=(\pa_{\lambda}B_{\mu\nu} + \pa_{\mu}B_{\nu\lambda}
+ \pa_{\nu}B_{\lambda\mu}). \label{h}
\ee
The only non-zero independent component of $\hi$ is
\be
H_{uij}=-(F_{ij} - {\bar F}_{ij}). \label{hu}
\ee

We now proceed to obtain the background field equations at tree level.
As mentioned earlier, these are provided by the one loop beta functions
of the couplings $G_{\mu\nu}$, $B_{\mu\nu}$ and $\phi$. 
Using standard expressions
for these equations \cite{sigmarev}, 
we obtain three independent equations of motion\cite{tsetrev}:
\be \pa^jF_{ij}=0, \hskip 1.5cm \pa^j{\bar F}_{ij}=0, \label{be1}
\ee
and
\be
- \hf\pa_i\pa^iK + \pa_i\pa^uA_{i}^{+} 
+ F_{ij}{\bar F}^{ij} + 2 \pa_{u}^{2}\phi
=0. \label{be2}
\ee
After obtaining the background field equations of the general K-models,
we now specialize to the case where the field strengths $F_{ij}$ and 
${\bar F}_{ij}$ are constant in the transverse directions but arbitrary
functions of $u$. A subclass of these, namely the ones which 
are constant in $u$ as well, describe the propagation of
closed strings in uniform magnetic field backgrounds\cite{russ}.
They have been shown to be represented by exact conformal field theory
models. The heterotic generalizations of these models 
have also been found. The chiral couplings in this case are,
\be
A_{i}=-\hf F_{ij}(u)x^j,  \hskip 1.5cm 
{\bar{A}}_{i}=-\hf {\bar{F}}_{ij}(u)x^j. \label{aisp}
\ee
As a consequence, we have for the curvature tensor, $R_{uijk}=0$ and
\be
R_{uiuj}= \hf \pa_i \pa_j K 
- {1\over {4}} G^{mn}(F_{jm} + {\bar F}_{jm})
(F_{in} + {\bar F}_{in}) 
- {1\over 2} \pa_u (F_{ij}(u) + {\bar{F}}_{ij}(u)) \label{rsp}.
\ee
In this case, the first two background field equations
in (\ref{be1}) are trivially satisfied.
For the metric equation we have,
\be
- \hf \pa^i\pa_i K(u,x) 
+ F_{ij}{\bar F}^{ij}(u) + 2 \pa_u^2 \phi (u) 
- {1\over 2} \pa_u (F_{ii}(u) + {\bar{F}}_{ii}(u)) =0. \label{besp}
\ee
Therefore $K$ must be a quadratic function of the $x^i$ in order to
satisfy equation (\ref{besp}).

We now proceed to show that this background 
is exact to all orders in 
$\alpha^{\prime}$. For this, we first note that 
that equation (\ref{besp})
is a second rank tensor equation. So we must consider all possible 
higher order second rank tensor contributions obtained from the background
field configuration. The only possible covariant tensor
components available for this purpose are, $D_u \phi$, $R_{uiuj}$ and its
covariant derivatives with respect to $D_u$ and $D_k$, 
$H_{uij}$ and its covariant
derivative with respect to $D_u$. 
One also has the corresponding contravariant
components which are consistent with the form of the metric (\ref{kmet}).

We first examine the terms involving a single Riemann tensor with the 
structure $D^{\lambda}D^{\nu}\rci$. An explicit evaluation provides
the identitites
\be
D^uD^u R_{uiuj}= D^iD^j R_{uiju}=D^uD^i R_{ujiu}=
D^iD^u R_{iuuj}=0 \;\;\;\; etc.,
\label{dr}
\ee
which implies
\be
D^{\lambda}D^{\nu}\rci=0 \label{drt}.
\ee
It is apparent now that to construct second 
rank tensors with $R_{uiuj}$ and its derivatives,
it is required to contract at least 
two indices of $R$ with another $R$
or its appropriate derivatives. Potentially 
non-zero contributions may 
come from the contractions of covariant indices $(u,i)$ and contravariant
indices $(v,i)$. However this requires a 
covariant index $v$ or contravariant
index $u$, which are unavailable and contractions 
on derivatives have been 
shown to be zero. Hence we conclude that it is impossible to construct
non-zero second rank tensors from contractions of $\rci$ and its derivatives
\cite{duval,asg}.
Thus all such higher order contributions are vanishing. 
Similarly notice that
terms of the form $D_{\lambda}\phi R^{\lambda\mu\nu\kappa}$ requires 
contraction
of covariant index $u$ and hence it is also zero.

We proceed to consider higher order contributions from the field strength
$H$ and its derivatives. Notice that the only non-zero component of $H$
is $H_{uij}$ and the only non-zero covariant derivative is $D_{u}H_{uij}$.
It is obvious that all terms involving only derivatives of 
$H$ requires contraction
of the covariant index $u$ which, as we showed earlier, 
was not possible. Hence these terms are all identically zero.
Higher order contributions of the schematic form 
$(DR)H$ and $(D\phi) H$ may be 
proved to be identically zero from similar considerations. It is also
possible to show that all scalars constructed from these covariant objects 
are also vansihing. Hence we conclude that the string background obtained 
from the K-models are exact to all orders in $\alpha^{\prime}$. In the 
next section we show how these backgrounds may be considered to be trivially
embedded in a type IIB string background and generate non-trivial type IIB 
backgrounds involving R-R fields.

\bigskip

\noindent {\bf 3. Type IIB Backgrounds and Ramond-Ramond Fields.}

\noindent In this section we now proceed to first show how the backgrounds
defined in eqns. (\ref{kmet}), (\ref{kb}) and $\phi$ 
can be embedded in a type IIB string theory with
vanishing R-R fields. We
subsequently present the action of the global $SL(2,R)$ transformations
on a type IIB background. Utilising these transfromations, we then
generate a non-trivial type IIB background involving R-R fields.
The field content of a type IIB string background consists of the following,
the string frame metric $G_{\mu \nu}$, two 3-form field
strengths $\hi^{(k)}$ where $k= (1, 2)$ , two scalars $\chi$ and $\phi$
from the NS-NS and R-R sectors respectively and a 5-form field strength
$F_{\lambda \mu \nu \kappa \rho}$. The two scalars $\chi$ and $\phi$
may be combined to form a complex scalar $\lambda =\chi + i e^{-\phi}$.
So we may consider the background obtained from the K-model
defined by $\phi (u)$, 
and equations (\ref{kmet}), (\ref{kb}) 
to be a special case of a type IIB background which has $\hi^{(2)}=0$
, $\chi=0$ and $F_{5}=0$. As shown in \cite{hull,berg}, 
type IIB strings in 
$D=10$ has a global $SL(2,R)$ symmetry at the level of the equations 
of motion \cite{berg,sch}. This acts on the type IIB 
background fields as follows :
\be
G_{\mu \nu}^{\prime}=\mid c\lambda + d \mid G_{\mu \nu}, \label{s1}
\ee

\be 
\lambda^{\prime}= {{a \lambda + b}\over {c \lambda + d}}, \label{s2}
\ee
and 
\be
\hi^{\prime (k)}=\Lambda \hi^{(k)}, \label{s3}
\ee
where $\Lambda$ is an $SL(2,R)$ matrix such that
\be
\Lambda = \left( \begin{array}{cc}
d & c\\
b & a\\
\end{array}\right) , \label{s4}
\ee
with $ad-bc=1$. 

Implementing the transformations outlined in eqns. (\ref{s1})-
(\ref{s3}),
we generate a nontrivial type IIB background
with R-R fields starting from the trivial type IIB configuration 
obtained from the K-models, {\it cf} eqns. 
(\ref{kmet}), (\ref{kb}) and $\phi(u)$.
Explicitly, we have
\be 
G_{\mu \nu}^{\prime}(u, x)=f(u) G_{\mu \nu}(u, x), \label{gp}
\ee
where $f(u)={\big [d^{2} + c^{2}e^{-2\phi (u)}\big ]}^{\hf}$
and
\be 
\lambda ^{\prime}={{iae^{-\phi} + b}\over 
{ice^{-\phi} + d}}, \label{lp}
\ee
with $\lambda^{\prime}=\chi^{\prime} + i e^{-\phi^{\prime}}$ and
$\lambda=i e^{-\phi}$.
We then have the final expressions for the type IIB scalars as:
\be
\chi^{\prime}(u)=
{1\over {{f(u)}^{2}}}\big [  db + ac\ e^{-2\phi}\big], \label{xp}
\ee
\be
\phi^{\prime}(u)=\phi (u) + 2\ ln\ f(u). \label{pp}
\ee
For
the 3-form field strength $H^{(k)}$, $k=1, 2$ we have
\be
\hi^{\prime (1)}=d \hi^{(1)}, \label{hp1}
\ee
and 
\be
\hi^{\prime (2)}=b \hi^{(1)}. \label{hp2}
\ee
The new metric is now given as:
\be
ds^{2}=2 f(u) dudv +f(u) dx^{i}dx_{i} + 2f(u)A_{i}^{+}dudx^i
+f(u) K (u,x) du^{2}, \label{nds}
\ee
where $K(u, x)=f(u)F(u, x)$. A rescaling $f(u)du=dU$ of the metric 
leads
to the general form 
\be
ds^{2}=-2 dUdv + {\tilde f}(U) dx^{i}dx_{i} 
+ 2 {\tilde A}_{i}^+ (U,x)dUdx^i
+ {{{\tilde K}(U,x)}\over{{\tilde f}(U)}}dU^{2}. \label{nds1}
\ee
Dropping the tildes and rewriting $U$ as $u$ in (\ref{nds1}) we have
\be
ds^{2}=2 dudv + 2 f(u)dx^{i}dx_{i} + 2 A_i^+(u,x)dudx^i
+ {\hat K}(u,x) du^{2}. \label{ndsf}
\ee
In subsequent discussions we drop the primes on the non-trivial type IIB
background fields generated by the $SL(2,R)$ transformations from the
K-model backgrounds. It can be seen from the definitions that 
$\hat{K}$ in equation (\ref{ndsf}) is also
a quadratic function of $x^i$'s. This fact becomes important in
proving that these backgrounds are all-order solutions of the
type IIB equations of motion.

As earlier, the $v$ independence leads to a null Killing vector $l^{\mu}$. 
We reexpress the metric in eqn (\ref{ndsf}) in terms of $l^\mu$ as
\be
G_{\mu\nu}=M_{\mu\nu} + {\hat K}l_\mu l_\nu. \label{gs}
\ee
where $M_{\mu\nu}$ is once again a $10 \times 10$ matrix given as
\be
M_{\mu\nu}=\left ( \begin{array} {ccc}
0 & 1 & A^{+}_{i}\\
1 & 0 & 0\\
A_{i}^{+} & 0 & f(u)I_{8}\\
\end{array} \right ) \label{ms}
\ee
and $I_8$ is a $8 \times 8$ unit matrix. The inverse metric may be
easily computed to obtain,
\be
G^{\mu\nu}=\left ( \begin{array}{ccc}
0 & 1 & 0\\
1 & -{\hat K} +{{A^{+2}_{i}}\over f} & -{{A_{i}^{+}}\over f}\\
0 & -{{A_{i}^{+}}\over f} & {{I_{8}}\over f}\\
\end{array}\right ) \label{gis}
\ee
Using these, it may be shown
that the only non-zero components of the Christoffel 
connections are, once again, 
$\gam^{v}_{uu}$, $\gam^{i}_{uu}$, $\gam^{v}_{ui}$,
$\gam^{j}_{ui}$ and $\gam^{v}_{ij}$. This leads to 
the null killing vector being covariantly constant i.e. 
$D_{\mu}l^{\nu}=0$ and $D_{\mu}l_{\nu}=0$. 

We now proceed to compute the Riemann curvature tensor for the 
metric (\ref{ndsf}) of the type IIB background generated by us.
Employing the closed form expression for the new metric, we
once again get the the Riemann tensor to be of the form in 
equation (\ref{rci})
with $\rci^{(M)}$ now being the Riemann tensor for the metric 
$M_{\mu \nu}$ in eqn. (\ref{ms}). 
Once again the only non-zero independent component
of the Riemann tensor is $R_{uiuj}$ when we specialize to the background
described by eqns. (\ref{aisp}). Explicitly, 
the expression for the Riemann tensor is,
\bea
R_{uiuj}& = &\hf \pa_i\pa_j K - \hf \pa_{u}\big [ \pa^iA_j^+ + 
\pa_jA_i^+ \big ] + {1\over 2} \pa_u^2 f \delta_{ij} \nonumber\\
&-& {1\over{4}}f^{-1}(u)\big [ -\pa_{u}f \delta_{ik} +
 ( \pa_i A_k^+ - \pa_k A_i^+)\big ] \times 
\big [ -\pa_{u}f \delta_{jk} + 
 ( \pa_jA_k^+ - \pa_k A_j^+)\big ]. \label{rusp}
\eea
The other background fields are the two 
antisymmetric tensor field strengths
from the NS-NS and the R-R sectors respectively, 
which are given by
$\hi^{(k)}$ and $(k=1, 2)$ in eqns (\ref{hp1}), (\ref{hp2}). 
Having obtained the type IIB background with 
non-trivial R-R fields, described
by eqns. (\ref{xp})-(\ref{hp2}) and (\ref{ndsf}), 
we now proceed to show that these are 
exact to all orders ( in $\alpha^{\prime}$).
We once again adopt the geometrical approach \cite{duval,asg} outlined
in section-2 for this purpose. The
background field equations for the type IIB superstring are, to the 
lowest order, those of N=2 D=10 supergravity in \cite{sch}.
They are all tensor equations of a definite rank .
For the type IIB background under 
consideration, we have second rank tensor
equations for the string form metric $G_{\mu\nu}$ 
and the antisymmetric tensor field $B_{\mu\nu}$.
We also have scalar equations 
for the NS-NS and the R-R scalars $\phi$, $\chi$
and a fifth rank completely antisymmetric
tensor equation for the five-form field strength 
$F_{\mu\nu\rho\sigma\kappa}$. The last equation
expresses the self duality condition on the five-form field strength. 

To study the all order contributions to the background field 
equations of motion, the possible corrections to all these tensor equations
must be considered. Notice that the contributions from the background
gauge fields $B_{\mu\nu}$ and $D_{\mu\nu\rho\sigma}$, appear in the higher
order terms as the corresponding gauge invariant field strengths. 
As a consequence, we need to consider the higher 
order terms in these equations
obtained from the following quantities, $R_{\mu\nu\rho\sigma}$, 
$H^{(k)}_{\mu\nu\rho}$,
$D_{\mu}\phi$, $D_{\mu}\chi$, $F_{\mu\nu\rho\sigma\kappa}$ and their 
covariant derivatives. In our case, we choose $F$ to be zero,
which is obtained by setting the four-form field
$D_{\mu\nu\rho\sigma}=0$ in its definition, together with the
form of $H^{(k)}$ in eqn. (\ref{hp1}), (\ref{hp2}). 

As in the bosonic case,
possible non-zero independent tensor components for the background
defined by eqns. (\ref{xp})-(\ref{hp2}) and (\ref{ndsf}) are 
$D_{u}\phi$, $ D_{u}\chi$, $R_{uiuj}$ and its
covariant derivatives with respect to
$D_{u}$ and $D_j$, $H_{uij}$ and its covariant
derivatives with respect to $D_{u}$ and 
the corresponding covariant components.
Notice that the index structures of the appropriate non-zero
tensor components are exactly as earlier in the trivial case.
The only additions for the field content in the 
non-trivial type IIB case are
the non-zero R-R fields $\chi (u)$ and
$H^{(2)}$ as the five-form field strength is zero. Hence
similar arguments show that all such higher order contributions
as earlier, are vanishing. For the two additional equations also,
namely scalar and the five-form R-R fields, 
similar geometrical arguments show that the higher
order constributions vanish. Hence the background field
equations which to the lowest order are those of N=2 D=10 supergravity,
are exact, to all-orders in $\alpha'$, 
also in presence of R-R fields in ten dimensions.

\bigskip
\noindent {\bf 4. Conclusions.}

\noindent To conclude, we have obtained a class of type IIB
superstring backgrounds involving R-R fields from the bosonic
K-models embedded in a
type IIB background with vanishing R-R fields. The complications
involving the sigma model couplings of the R-R fields have been
obviated by adopting a purely geometrical approach to
compute the higher order terms in the equations of motion. 
This approach,
which is based on the analysis of the tensorial 
index structure of higher
order contributions, seems to be applicable to a wider
class of backgrounds than those obtained from the K-models.
We mention {\it en passant} that our analysis has been restricted
to K-model backgrounds with vanishing
five form self-dual field strength. Furthermore we have focussed
on strictly $u$ dependent
antisymmetric tensor field strengths and dilaton. 
It would be an interesting exercise to obtain type IIB 
backgrounds from more general K-models and show that they are
also exact following the geometrical approach which has been
elucidated in this article. 
In particular, for K-models in the bosonic case, 
$A=0$ or $\bar{A} =0$ conditions 
provide another class of all-order
solutions. It will be interesting to show that they are exact
for type IIB as well. 

Our results also indicate that type IIB strings in a constant 
magnetic field background may be formulated as an exact
Conformal field theory. It will be interesting to study the
phase transition, discussed in \cite{russ}, for this case.
The status of unbroken space-time supersymmetries, like the
K-models with trivial embeddings in superstrings, is also of
interest to investigate in presence of R-R fields. These will
have implications for the present backgrounds to be solutions in
the presence of local string-loop corrections as well.

\noindent {\bf Acknowledgements:} G.S would like to thank Institute of
Physics, Bhubaneswar where this work was partially completed
and the High Energy Theory Group there for the warm hospitaliy
and the stimulating research environment.
\vfil\eject

\end{document}